\documentstyle[12pt,aas2pp4]{article}

\def\alwaysmath#1{\ifmmode{#1}\else{$#1$}\fi}

\def\morethan#1{{ et al.}}

\slugcomment{\footnotesize In press to ApJ.
}

\lefthead{Ventura et al.}
\righthead{Lithium rich carbon stars in the Magellanic Clouds}

\begin{document}

 \title{Modeling lithium rich carbon stars in the Large Magellanic
 Cloud: an independent distance indicator ?} 

\author{Paolo Ventura \altaffilmark{1},  
Francesca D'Antona \altaffilmark{1},  
Italo Mazzitelli \altaffilmark{2}
}
\altaffiltext{1}{Osservatorio Astronomico di Roma, Via Frascati 33, 
00040 Monte Porzio Catone,(Rome), Italy}
\altaffiltext{2}{Istituto di Astrofisica Spaziale del CNR, Via del Fosso
del Cavaliere 100, 00133  Rome, Italy 
}

\begin{abstract}

We present the first quantitative computations explaining the presence in the
Large Magellanic Cloud of some AGB stars which share the properties of
being Carbon stars (surface C/O$>1$) and Lithium rich.
A self--consistent description of
time--dependent mixing, overshooting and nuclear burning was required.
The products of nucleosynthesis at the stellar surface turn out to be very
sensitive to the temperature at the base of the outer convective envelope
($T_{bce}$) during the quiescent phase of hydrogen burning. Lithium
production is obtained for $T_{bce}\geq 4\cdot 10^7 K$ (Hot Bottom Burning),
but $T_{bce}\geq 6.5\cdot 10^7 K$ is necessary to cycle into Nitrogen 
the carbon
previously convected to the stellar surface by the third dredge up.
Therefore, Li--rich C stars can occurr for $T_{bce}$'s in this small range of
temperatures. We then identify a possible -narrow- range of masses and {\it
luminosities} for this peculiar evolutionary stage. Comparison of these
models with the luminosities of the few Li--rich C stars in the LMC provides
an independent distance indicator (within $\sim 0.25$mag) for the LMC.
Present data and models are consistent with $(m-M)_0 \sim 18.7$, but a better
determination would be possible by refining the observations and the theoretical
models. 

\end{abstract}

\keywords{stars: evolution, AGB and post-AGB, carbon}

\section{Introduction}

Spectroscopic analysis in the Magellanic Clouds (Smith \& Lambert
1989, 1990) showed the presence of high--luminosity, lithium rich AGB stars,
having surface Li abundances orders of magnitude larger than commonly observed
in galactic giants. The presence of Li is well explained in the framework of
``hot bottom burning" (HBB), as suggested by Cameron and Fowler (1971).
Models coping with this problem must employ non instantaneous mixing:
in fact, the Cameron--Fowler mechanism is based on a nuclear
time scale for Li and Be--burning shorter than that of mixing
(Sackmann \& Boothroyd 1992).
Our code, ATON2.0, has been recently implemented with an algorithm coupling
nuclear burning and time--dependent chemical mixing (and overshooting, if
any), allowing it to successfully deal with the above problem (Ventura et al.
1998; Mazzitelli, D'Antona \& Ventura 1999).

In this paper we focus our attention on Carbon stars (C stars) in the
LMC that exhibit also large Li abundances at their surfaces (Richer, Olander
\& Westerlund 1979, Smith, Plez \& Lambert 1995), since the co--existence 
of C and Li is unexpected
on theoretical grounds. In fact, the Hot Bottom Burning (HBB)
conditions required to ignite Li production should lead to fast CNO
processing at the base of the envelope, with subsequent 
destruction of the $^{12}C$
previously convected to the surface during the third dredge--up. We show that
there is a narrow range of stellar masses in which, despite the activation of
the Cameron--Fowler mechanism, carbon can survive long enough in the
convective envelope to keep the C/O ratio over unity, although the
$^{13}$C/$^{12}$C ratio increases. For a non negligible
fraction of their AGB life, these objects can be then observed as Li--rich
C stars.

\section{The Models}
The main features of the ATON2.0 code are described in Ventura et al.
(1998). Here we only mention the main updates in the macrophysical inputs.
Even if some convective overshooting is included in the present models, 
convection as a whole is addressed in a purely local
frame. The formal convective boundaries are thus fixed according to the 
Schwarzschild criterion. The convective fluxes inside the
instability regions follow the full spectrum of turbulence (FST) model of
convection, in the form given by Canuto, Goldman \& Mazzitelli (1996), 
to account for the non--linearities in the growth rates.
Fine tuning of the $\beta$ parameter in FST model in not necessary 
in this 
context, since we know that also a variation by $50\%$ does not show any
important effect upon lithium production (Mazzitelli et al. 1999),
at variance with the results obtained by the mixing length
theory (MLT) (Sackmann \& Boothroyd 1992, 1995).

Lithium production can be modeled only by 
adopting a time--dependent approach, in which nuclear
burning and mixing of chemicals are self--consistently coupled.
In our code, this goal is accomplished by solving a
diffusive equation that is valid for any chemical species. In our algorithm, 
the terms relative to nuclear evolution and mixing
are treated simultaneously: the approximation of dealing separately with 
each of them would lead, in the present case, to a vanishingly small
abundance of Li (Mazzitelli et al. 1999).

Our scheme of overshooting from the convective regions (Ventura et al. 1998) is
based on an exponential decay of the mean turbulent velocity outside the
convective boundaries, in the form: $u=u_b\cdot exp ({1\over \zeta
f_{thick}}ln{P\over P_b})$ \, where $u_b$ and $P_b$ are the velocity and
pressure at the convective boundary, $P$ is the local pressure, and $\zeta$
is a free parameter measuring the extent of the overshooting region.
Moreover, $f_{thick}$ is the thickness of the convective region in units
of the pressure scale heigth $H_p$\ (maximum allowed value = 1), and this
avoids overshooting regions larger than the entire convective zone.
This treatment of overshooting is not much different from the one adopted by
Herwig et al. (1997); it is also consistent with the results of numerical
two--dimensional hydrodynamic simulations by Freytag, Ludwig \&
Steffen (1996).
In the present work, being interested to investigate the properties of
carbon stars, we assumed both overshooting from above and below convective
regions, and the value of the parameter $\zeta$ has been set equal to 0.02,
in agreement with previous works.  

\section{Lithium production and carbon stars}

The most luminous AGB stars in the Magellanic Clouds are
Li--rich (Smith \& Lambert 1989, 1990), consistent
with the mechanism of Li production by HBB hypotesized by Cameron \& Fowler
(1971). For $T\geq 4\cdot 10^7K$, the reaction $^3He+^4He \longrightarrow
^7Be$\ becomes efficient on the timescale of AGB evolution.
Beryllium can then capture one more proton or decay into lithium.
In this latter case, the mixing time scale is such that, in a dynamical
equilibrium, some lithium is produced far from the base of the envelope.
Here it survives due to the low temperatures and can eventually reach the surface.
Independently of the above process, carbon stars form as a
consequence of the penetration of the convective envelope into inner layers,
following each thermal pulse (TP). Primary carbon is dredged up to the
surface until the C/O ratio grows larger than unity. C stars are usually
observed at lower luminosities than Li--rich AGBs, since, for large
temperatures at the base of the convective envelope, the CNO cycle becomes so
efficient that carbon is turned into nitrogen faster than it is dredged up.

This is suggested both by computations and
by observations showing that practically all C stars are less luminous than
Li--rich stars in the Magellanic Clouds. A recent survey by
Plez, Smith \& Lambert (1993) has however confirmed the presence of
Li--rich C stars. Also in our Galaxy a few
Li--rich C stars are known (Wallerstein \& Conti 1969; Abia et al. 1991):
they seem to be among the most luminous C stars, but their distance
determinations are not so certain. On the other hand, some Magellanic Clouds
Li--rich C stars are certainly among the most luminous C stars, just below
the O--rich, Li--rich AGB stars (Smith et al. 1995).

Our aim was then to investigate a theoretical framework in which the coexistence
of C and Li--rich stars is allowed.
We focused our attention on masses $M \leq 4M_\odot$.
The tracks were computed with a  metallicity appropiate to the
Large Magellanic Cloud, namely Z=0.01. Mass loss has been modelled according to
Bl\"ocker (1995), with $\dot M= \dot M_R \cdot L^{2.7}/M^{2.1}$, being $\dot
M_R$ the usual Reimer's mass loss rate. In this latter, the free parameter
$\eta$ has been set to 0.05, since this value seems to fit the lithium
vs. luminosity trend observed in the Magellanic Clouds AGB stars (Ventura,
D'Antona \& Mazzitelli 1999).


\begin{figure}[h]
\caption{Variation with time of the physical and chemical properties
of a $4M_{\odot}$ model of metallicity $Z=0.01$ during the AGB phase of
thermal pulses. Left panels show the evolution of luminosity and of the
temperature at the base of the convective envelope, right side panels report
the changes in surface Li and  C/O.}
\label{fig_1}
\end{figure}

Fig.1 shows the physical and chemical evolution of a $4M_{\odot}$ sequence.
The trend of the run luminosity vs. time after the first five pulses, is
different from the case when overshooting from below the bottom of the
convective envelope is neglected. The explanation is in the deeper diffusion
of chemicals, which mixes up in the envelope helium produced by the
H--burning shell. A detailed analysis of the mechanism, and of the
differences encountered on the AGB when applying the two different schemes of complete,
instantaneous overshooting, and the diffusive, time--dependent one, can be found
in Mazzitelli et al. (1999).

Let us now examine the evolution of chemicals at the surface of the star. We
first see an increase in the C/O ratio following each pulse, due to an early
onset of the third dredge--up. At the same time, the Cameron--Fowler
mechanism ignites and the lithium abundance, which had previously dropped to
$\log[\epsilon(^7Li)]\sim -2$ (where $\log[\epsilon(^7Li)]=\log(^7Li/H)+12)$,
starts rising. At the third pulse the star
already shows up as Li--rich, but it never becomes a C star since CNO
burning grows more and more efficient at the base of the envelope. This is
seen in the right lower panel of fig.1, where a severe drop of the C/O ratio
following the H--shell re--ignition after each pulse is evident.
From these results, we conclude that C stars can not be produced or survive
when $T_{bce}\geq 6.5\cdot 10^7 K$. At variance, 
the Cameron--Fowler mechanism starts leading to detectable
surface lithium abundances already at $T_{bce}\sim 4\cdot 10^7 K$.
There is then a narrow range of $\delta T_{bce}
\sim 2.5\cdot 10^7$ K in which, in principle, both C and Li can
show up at the surface, if the third dredge--up is already efficient.
It is then likely that the observed 
Li--rich C stars in the Magellanic Clouds have $T_{bce}$'s
falling in this range. 
 
Consider now the evolution of smaller masses, namely M/M$_{\odot}=3.8$,
3.5 and 3.3, for the same metallicity. The results are shown in fig. 2.
The $3.8M_{\odot}$ track starts showing large lithium abundances shortly
after the C/O ratio exceeds unity but, after one more TP, hot bottom burning
becomes effective in depleting carbon. 
The ratio $^{13}$C/$^{12}$C during the evolution rises to $\sim 0.4$, thus this
object has the characteristics of a J--type star.
From fig.2, one can conclude that the
star is a Li--rich C star for $\sim 30000$ years.

\begin{figure}[h]
\caption{Variation with time of the luminosity, $T_{bce}$, Li abundance and
C/O ratio in three intermediate mass models of different masses during the
phase of thermal pulses.}
\label{fig_2}
\end{figure}


The track for the $3.5M_{\odot}$ star is the most suitable to fit the
observational constraints of being both C-- and Li--rich. $T_{bce}$\ rapidly
grows hot enough to ignite lithium production but, before the CNO cycle
takes over, one has to wait for $\sim 2 \cdot 10^5$ yr. Moreover, shortly
after the ignition of CNO burning at the base of the envelope, mass loss is
so efficient that it works as a thermostat, and the C/O ratio does not
decrease any more. This is confirmed by the drop of the total luminosity and
of the temperature at the base of the convective envelope of the star shown
in fig.2. The track was stopped when the total mass of the star was slightly
lower than $1.4M_{\odot}$\ during which time lithium was also decreasing. Here again the
$^{13}$C/$^{12}$C ratio is $\sim 0.24 - 0.3$ during the Li--rich phase.

Lastly, we discuss the track for the $3.3M_{\odot}$ star. This case looks
similar to the Li--production mechanism discussed by Iben (1973).
$T_{bce}$\ never grows large enough to ignite the Cameron--Fowler mechanism,
and the star never produces lithium, except for short periods during TP's:
following the peak of the pulse, the external envelope penetrates
inwards, and convective eddies reach sufficiently hot regions 
to ignite $^3$He burning, which in turn 
triggers beryllium production, which then decays into lithium in the envelope.
At the same time more $^{12}$C is convected to the surface of the star.
Afterwards, lithium production ends, buth lithium survives then in the
envelope until the H--burning shell resumes, $T_{bce}$ grows 
again to $\geq 10^7K$,
a temperature not sufficient to provide $^3$He burning, but large enough to 
destroy lithium.
For lower masses,
not only is lithium more severely depleted in pre--AGB phases, but the
efficiency of the third dredge--up does not allow for both C-- and
Li--rich structures. The evolution shows no sign of a $^{13}$C/$^{12}$C increase.

The above results illustrate the following:\\
1) there might be (rarely) Li--rich C stars in the middle of the O-rich
Li--rich AGBs which populate the $M_{bol}$\ range between -6 and -7. In
particular, the 3.8$M_\odot$\ evolution presented here shows that for a
relatively short lifetime this mass could show up as a Li--rich C star 
also rich in $^{13}$C at
$\log L/L_\odot \simeq 4.5$, that is at $M_{bol}=-6.5$.
Incidentally we note that Crabtree, Richer \& Westerlund (1976) have discovered in the LMC
such a luminous atypical Li--rich C star.

2) stars such as the $3.5M_{\odot}$\ one present quite a long evolutionary stage,
in which they can be observed as Li--rich C stars. Their luminosity should be
$\log L/L_\odot \simeq 4.2 - 4.28$\ (namely $M_{bol} = -5.75 - -5.95$). 
The most
luminous Li--rich C stars {\it below} the boundary of $M_{bol}=-6$\ observed
in the LMC are then likely to fall within this range; 
they should also be
J stars, being very rich in $^{13}$C. There are three such
objects, listed in the Smith et al. 1995 paper, with $M_{bol}$\ 
ranging from -5.5
to -5.7, based on a distance modulus $(m-M)_0=18.6$.
If we take these numbers at face value, a distance modulus $(m-M)_0 \sim 18.7$
would give the best match between our models and the observations. However,
the errors on the observed absolute bolometric magnitudes are certainly
too large (at least $\pm 0.2 - 0.3$mag according to Smith et al. 1995) to draw
strong conclusions. Nevertheless, the range of masses 
and luminosities for which the HBB conditions required to produce 
Li--rich C stars are fullfilled is so
small that, given sufficiently accurate theoretical models
(see discussion in section 4),
{\it a better determination 
of the observational magnitudes, 
and possibly an enlarged sample of 
such stars could constitute a powerful,
independent way of determining the LMC distance, 
an important intermediate 
benchmark to the distance scale of the Universe}. A similar suggestion was
advanced earlier by Plez et al. (1993), 
who proposed to use the O--rich Li--rich
AGBs having $M_{bol}$\ between -6 and -7. The present suggestion may lead to
a more accurate determination of distance, in view of the smaller range of
magnitudes of these stars.

3) In the context of our modeling it is {\it not} possible to give a
satisfactory explanation for the Li--rich C stars of $M_{bol} \sim -4.5$
($\log(L/L_{\odot}) \sim 3.8$). 
There are three such stars in Smith et al. (1995),
and three also in the M31 survey of C stars by Brewer, Richer \&
Crabtree (1996), and all 
of them are J stars. If these stars belong to the evolution of the 
$3.5M_{\odot}$ stars, they can be at such a low luminosity only for $\sim 6 \%$
of the time (during the lowest luminosity phase of the TPs). In this case,
there should be $\sim 15$ Li--rich C stars at $M_{bol}\sim -5.7$ for each
star at $M_{bol} \sim -4.6$. These stars are just not observed and would not
have been missed in the quoted surveys.

On the other hand, stars following the 3.3$M_\odot$\ evolution could show 
up as Li--rich C stars
for short periods. However, these stars should {\it not} be J stars, as
$^{13}$C is not enhanced in our computations. We leave this problem open. 

\section{The theoretical luminosity range of Li--rich C stars}
To first order, the luminosity at which the $T_{bce}$ needed to ignite 
lithium production is reached is not very dependent 
on the convective model adopted, 
as shown by Mazzitelli et al. (1999) for Li--rich O--rich AGB modeling.
The effect of changing the convection model and overshooting is mainly to
shift the mass range over which Li--rich O--rich stars appear, but the 
luminosity range at which this occurs is very similar:
we have verified this coincidence within $\sim 0.1$ mag.
Also, the above quantities do not vary if, in modeling overshooting,
we change the e--folding distance of decay of velocity outside the 
convective regions (consistently with early onset of the third dredge--up).
The same reasoning can be applied to the Li--rich C stars. Neverthless, 
however, a wide exploration of physical parameters (chemical composition,
opacities, convective model, mass loss rate) would be needed to 
determine the theoretical dependence of the luminosity range of Li--rich
C stars to better than 0.1 mag. Of course, we did not take into account
effects such as the difference between upward and downward convective flows,
eddy transport over scales large compared to the scale of diffusive mixing
assumed in the FST theory, etc.
In practice, from a theoretical point of view,
much more work is needed to allow use of Li--rich C stars as a good
distance indicator.

\section{Conclusions}
In this paper, we present and discuss the first truly evolutionary models
successfully reproducing the status of Li--rich C stars.
We find that this modeling is made possible by a small difference
between the temperatures required to ignite the Cameron--Fowler mechanism
leading to lithium production, namely $T_{bce}\sim 4\cdot 10^7K$,
and the one required to destroy carbon by CNO fusion, that is:
$T_{bce}\sim 6.5\cdot 10^7K$. We conclude that the Li--rich C stars
in the Magellanic Clouds have temperatures at the base of their convective 
envelopes within the range identified by the two above values. 
Although much wider modeling will be needed to check the uncertainties, 
this result
is at first order independent of some of the macro--physical input, like 
the convective model and the extent of the overshooting region.
{\it The range of luminosities in which this process may take place is very
small: $\log L/L_\odot \simeq 4.2 - 4.28$, namely $M_{bol} \simeq -5.75 -
-5.95$}: thus the luminosity of Li--rich C stars can become an interesting
independent method of determining the LMC distance modulus if we succeed
in refining the observations and the theoretical models.

\acknowledgments
We thank the anonymous referee for a careful report and H.Richer for
useful suggestions.

\end{document}